\documentclass{birkmult}

\theoremstyle{definition} 

\theoremstyle{remark}

\numberwithin{equation}{section}

\begin{document}

\title[New Types of Solutions of NLFDE]
{New Types of Solutions of Non-Linear \\Fractional Differential Equations}

\author[Edelman]{Mark Edelman}

\address{%
Stern College for Women at Yeshiva University\\
245 Lexington Ave.\\
New York, NY 10016\\
USA}

\email{edelman@cims.nyu.edu}

\thanks{This work was completed with the support of Yeshiva University,
Courant Institute of Mathematical Sciences, and DOE Grant DE-FG0286ER53223}
\author{Laura Anna Taieb}
\address{%
Stern College for Women at Yeshiva University\\
245 Lexington Ave.\\
New York, NY 10016\\
USA}

\email{taieb@yu.edu}


\date{July 12, 2011}

\begin{abstract}
Using the Riemann-Liouville and Caputo Fractional Standard Maps (FSM) 
and the Fractional Dissipative Standard Map (FDSM) as examples, 
we investigate types of solutions of non-linear
fractional differential equations. They include
periodic sinks, attracting slow diverging trajectories (ASDT),
attracting accelerator mode trajectories (AMT), 
chaotic attractors, and cascade of
bifurcations type trajectories (CBTT). New features discovered include
attractors which overlap, trajectories which intersect, and CBTTs.
\end{abstract}

\maketitle

\section{Introduction}
\label{int}

In recent years fractional calculus (FC) and fractional differential
equations (FDE) became very popular in many areas of science. The books
dedicated to the applications of FDEs published in 2010-2011 include 
\cite{TarasovBook2011,TarasovBook2010,HerrmannBook2011} in physics in general, 
\cite{ControlBook2010,ControlBook2011} in modeling and control, 
\cite{MainardiBook2010} in
viscoelasticity, \cite{LuoBook2010} in systems with long-range interaction.
A good review of applications of FC to chaos in Hamiltonian Systems is given
in \cite{ZaslavskyBook2005}. Because fractional derivatives are
integro-differential operators, they are used to describe systems
distributed in time and/or space: systems with long range interaction 
\cite{LuoBook2010,LZ2006,TZ2006,KZ2007,ZET2007,TZ2007}, non-Markovian systems
with memory (\cite{Podlubny} Ch.10, \cite{Mem1,Mem2,Mem3,Mem4,Mem5}),
fractal media \cite{Mainardi}, etc. Biological systems are probably the best
examples of systems with memory. As it has been shown recently 
\cite{Lund1,Lund2}, even processing of external stimuli by individual neurons
can be described by fractional differentiation. In some cases 
\cite{KST,KBT1,KBT2,TarV} FDEs are equivalent 
to the Volterra integral equations
of the second kind. This kind of equations (non necessarily FDEs) is used in
nonlinear viscoelasticity (see for example \cite{Wineman2007,Wineman2009})
and in biology (for applications to mathematical models in population
biology and epidemiology see \cite{HoppBook1975,PopBioBook2001}).

As in regular dynamics (in application to various areas in science), the
non-linearity plays a significant role in fractional dynamics. Chaos and
order in nonlinear systems with long range interactions were considered in 
\cite{LuoBook2010,TZ2006,KZ2007,ZET2007}, nonlinear FDEs in application to
the control were reviewed in books \cite{ControlBook2010,ControlBook2011},
nonlinear fractional reaction-diffusion systems considered in \cite%
{Gafiychuk2010,Guf1,Guf2,Guf3}. Corresponding to the fact that in physical
systems the transition from integer order time derivatives to fractional (of
a lesser order) introduces additional damping similar in appearance to
additional friction \cite{Podlubny,ZSE}, phase spaces of the systems with
fractional time derivatives demonstrate different kinds of structures
similar to the attractors of the dissipative dynamical systems \cite%
{ControlBook2010,ZSE,TH}.

As in the case of the systems which can be described by regular differential
equations, general properties of systems described by FDEs can be
demonstrated on the examples of maps which can be derived by integrating the
FDEs (if it is possible) over a period of perturbation. Equations of some
fractional maps (FM), which include among others fractional standard maps
(FSM) and fractional dissipative standard
maps (FDSM), were derived in a series of
recent publications \cite{LuoBook2010,TZ1,ETFSM,TEdisFM,TarMap1,TarV} 
from the corresponding FDEs.  

Two of the most studied  examples in the regular case are the Chirikov 
standard map (SM) \cite{Chirikov}
and the Zaslavsky dissipative standard map (DSM) \cite{DM1,DM2}. 
The SM provides the simplest model of the universal 
generic area preserving map 
and the description of many physical systems and 
effects (Fermi acceleration, comet dynamics, etc.) can be reduced   
to the studying of the SM. The topics examined include 
fixed points, elementary structures of islands and a chaotic sea,
and fractional kinetics \cite{Chirikov,LichLib,Zaslavsky1}. 
Different properties of the DSM were discussed in 
\cite{r2,r3,r5a,r5b,r5c,r8}, and a
rigorous proof of the existence of chaotic attractor in such type of
systems was obtained in \cite{r6,r7}. It was also proved in \cite{r6,r7} 
that the system of this type exhibits quasi-periodic attractors, 
periodic sinks,
transient chaos with the dynamics attracting to the sink, and the existence
of the SRB measure for the purely chaotic dynamics. 
All this stimulated the study of the 
FSMs \cite{ETFSM,myFSM}  and FDSM \cite{TEdisFM} 
in hope to reveal the most general properties of the fractional dynamics.

Two-dimensional maps investigated in  
\cite{ETFSM,TEdisFM,myFSM} were derived from the
FDEs (see Section \ref{BE}) and, as a result, 
they are discrete maps with memory
in which the present state of evolution depends on all past states.
The earlier studies of maps with memory were done on one-dimensional maps
which were not derived from differential equations
\cite{Ful,Fick1,Fick2,Giona,Gallas,Stan}.
Most results were obtained 
for the generalizations of the logistic map and the main general result
is that the presence of memory makes systems more stable. 
The initial study of the FSM 
in \cite{ETFSM,TEdisFM,myFSM} was concentrated on the investigation of
the fixed/periodic points' stability of the FSM,  new properties
and new types of attractors.

The results obtained in \cite{ETFSM,TEdisFM,myFSM} revealed new
unusual properties of the fractional attractors of the FSM, which are
different not  only from the properties of the 
fixed/periodic points of the non-dissipative systems (the SM in our case), 
but also from the
properties of attractors of the regular (not fractional) 
dissipative systems (e.g. the DSM). 
The most unusual observed property of the FSMs
is the existence (and persistence) of the new type of attractors - 
cascade of bifurcations type trajectories (CBTT). 
A cascade of bifurcations, when with the change in the value of a
parameter a system undergoes a sequence of period-doubling bifurcations,
is a well known pathway of a transition form order to chaos.
In the case of the CBTT the period doubling occurs without a change in a
system parameter and is an internal property of a system. 
In Section \ref{R} we summarize results of the extensive 
numerical investigation of three fractional maps
to demonstrate the properties of fractional attractors. An analysis of
the conditions for the CBTTs' appearance will be presented in the
following publications.

\section{Basic Equations}
\label{BE}

The standard map in the form

\begin{equation}  \label{SM} 
p_{n+1} = p_n - K \sin x_n, \ \  x_{n+1} = x_n + p_{n+1}  \  \ ({\rm mod} \ 2\pi )
\end{equation}
can be derived from the differential equation
\begin{equation}
\ddot{x}+K \sin(x) \sum^{\infty}_{n=0} \delta \Bigl(\frac{t}{T}-(n+\varepsilon) \Bigr)=0,
\label{SMDE}
\end{equation}
where $\varepsilon  \rightarrow 0+$, following the steps proposed  by
Tarasov in  \cite{TarV}.

The equations for the Riemann-Liouville FSM (FSMRL)
were obtained in \cite{TZ1} and \cite{TarV}.
Following the steps proposed \cite{TarV}, the FSMRL in the form which
converges to the SM as $\alpha  \rightarrow 2$ can be derived from the 
differential equation with
the Riemann-Liouville fractional derivative describing a kicked system
\begin{equation} \label{difFSMRL}
_0D^{\alpha}_t x+K\sin(x) \sum^{\infty}_{n=0} \delta \Bigl(\frac{t}{T}-(n+\varepsilon) \Bigr)=0, 
\quad (1 <\alpha \le 2), 
\end{equation}
where $\varepsilon  \rightarrow 0+$, with the initial conditions 
\begin{equation}
(_0D^{\alpha-1}_tx) (0+) = p_1, \   \
(_0D^{\alpha-2}_tx) (0+) = b  ,
\label{FSMRLic}
\end{equation}
where
$$
_0D^{\alpha}_t x(t)=D^n_t \ _0I^{n-\alpha}_t x(t)=
$$
\begin{equation}
\frac{1}{\Gamma(n-\alpha)} \frac{d^n}{dt^n} \int^{t}_0 
\frac{x(\tau) d \tau}{(t-\tau)^{\alpha-n+1}}  \quad (n-1 <\alpha \le n),
\label{RL}
\end{equation}
$D^n_t=d^n/dt^n$, and $ _0I^{\alpha}_t$ is a fractional integral.

After integration of equation (\ref{difFSMRL}) the 
FSMRL can be written in the form
\begin{equation} \label{FSMRLp}
p_{n+1} = p_n - K \sin x_n ,
\end{equation}
\begin{equation} \label{FSMRLx}
x_{n+1} = \frac{1}{\Gamma (\alpha )} 
\sum_{i=0}^{n} p_{i+1}V^1_{\alpha}(n-i+1) 
, \ \ \ \ ({\rm mod} \ 2\pi ) ,
\end{equation}
where 
\begin{equation} \label{V1}
V^k_{\alpha}(m)=m^{\alpha -k}-(m-1)^{\alpha -k} 
\end{equation}
and momentum $p(t)$ is defined as
\begin{equation} \label{MomRL}
p(t)= \, _0D^{\alpha-1}_t x(t).
\end{equation} 
Here it is assumed that $T=1$ and $1<\alpha\le2$. The condition $b=0$
is required in order to have solutions bounded at $t=0$ for
$\alpha<2$ \cite{ETFSM}. In this form the FSMRL equations in the limiting
case $\alpha=2$ coincide with the equations for the standard map under the
condition  $x_0=0$. For consistency and in order to compare corresponding
results for all three maps (the SM, the FSMRL, and the Caputo FSM (FSMC)) 
all trajectories considered in this article have the initial condition  $x_0=0$.

Following the steps proposed \cite{TarV}, the FSMC 
in the form which
converges to the SM as $\alpha  \rightarrow 2$ can be derived from the 
differential equation similar to (\ref{difFSMRL}) but with
the Caputo fractional derivative
\begin{equation} \label{difFSMC}
_0^CD^{\alpha}_t x+K\sin(x) \sum^{\infty}_{n=0} \delta \Bigl(\frac{t}{T}-(n+\varepsilon) \Bigr)=0, 
\quad (1 <\alpha \le 2) 
\end{equation}
where $\varepsilon  \rightarrow 0+$, with the initial conditions 
\begin{equation}
p(0)=(_0^CD^{1}_tx) (0) = (D^1_tx) (0)=p_0, \   \  x(0)=x_0  ,
\label{FSMCic}
\end{equation}
where
$$
_0^CD^{\alpha}_t x(t)=_0I^{n-\alpha}_t \ D^n_t x(t) =
$$
\begin{equation}
\frac{1}{\Gamma(n-\alpha)}  \int^{t}_0 
\frac{ D^n_{\tau}x(\tau) d \tau}{(t-\tau)^{\alpha-n+1}}  \quad (n-1 <\alpha \le n).
\label{Cap}
\end{equation}

Integrating equation (\ref{difFSMC}) with the momentum defined as
$p=\dot{x}$ and assuming $T=1$ and $1<\alpha\le2$, 
one can derive the FSMC in the form
$$
p_{n+1} = p_n 
-\frac{K}{\Gamma (\alpha -1 )} 
\Bigl[ \sum_{i=0}^{n-1} V^2_{\alpha}(n-i+1) \sin x_i 
$$
\begin{equation} \label{FSMCp}
+ \sin x_n \Bigr],\ \ ({\rm mod} \ 2\pi ), 
\end{equation}
$$
x_{n+1} = x_n + p_0 
$$
\begin{equation} \label{FSMCx}
-\frac{K}{\Gamma (\alpha)} 
\sum_{i=0}^{n} V^1_{\alpha}(n-i+1) \sin x_i,\ \ ({\rm mod} \ 2\pi ). 
\end{equation}
It is important to note that the FSMC ((\ref{FSMCp}),   (\ref{FSMCx})) can
be considered on a torus ($x$ and $p$ mod  $2 \pi$), a cylinder 
($x$  mod  $2 \pi$),  or in an unbounded phase space,
whereas the FSMRL ((\ref{FSMRLp}), (\ref{FSMRLx}))  
can be considered only in a cylindrical or
an unbounded phase space. The  FSMRL  has no periodicity in $p$  
and cannot be considered on a torus. This fact is related to the
definition of momentum (\ref{MomRL}) and initial conditions 
(\ref{FSMRLic}). The comparison of the phase portraits of  two FSMs
is still possible if we compare the values of the $x$ coordinates on
the trajectories corresponding to the same values of the maps' parameters.

The DSM in the form 
\begin{equation} \label{Zasl1}
X_{n+1}=X_n+P_{n+1} ,
\end{equation}
\begin{equation} \label{Zasl2}
P_{n+1}=-bP_n-Z \, \sin(X_n) 
\end{equation}
can be derived integrating the differential equation of the the kicked
damped rotator (see for example \cite{TEdisFM})
\begin{equation} \label{dr-0b}
\ddot{X}+ q \dot{X}=\varepsilon \sin(X) 
\sum^{\infty}_{n=0} \delta \Bigl(t -n \Bigr) .
\end{equation}

Two forms of the FDSM were derived in  \cite{LuoBook2010,TZ1,TEdisFM}. 
The form which has been investigated numerically in  \cite{TEdisFM}
\begin{equation}  \label{T1}
X_{n+1} = \frac{\mu^{-1}}{\Gamma(\alpha-1)} 
\sum^{n}_{k=0} P_{k+1} W_{\alpha}(q,k-n-1) ,
\end{equation}
\begin{equation}  \label{T2}
P_{n+1}= - b P_n -Z \sin ( X_n ) ,
\end{equation}
where functions $W_{\alpha}(a,b)$ are defined by 
\begin{equation}  \label{Wa}
W_{\alpha}(a,b)=a^{1-\alpha}  e^{a(b+1)} \,
\Bigl[ \Gamma(\alpha-1, ab ) - \Gamma(\alpha-1, a(b+1) ) \Bigr] 
\end{equation} 
and $\Gamma(a,b)$ is the incomplete Gamma function  
\begin{equation} \label{incomplete}
\Gamma(a,b)=\int^{\infty}_b y^{a-1} e^{-y} dy 
\end{equation}
has been derived from the following fractional generalization of (\ref{dr-0b}) 
\begin{equation} \label{fdr}
_0D^{\alpha}_t X(t) - q \, _0D^{\beta}_t X(t) =
\varepsilon \sin(X) \sum^{\infty}_{n=0} \delta (t -n ) ,
\end{equation}
where
\[ q \in \mathbb{R}, \quad 1< \alpha\le 2 , \quad \beta=\alpha-1 . \]
Here 
\begin{equation}
\mu = (1-e^{-q})/q , \quad Z = - \mu \varepsilon e^q . 
\end{equation}
Further in this paper we will also use
$K=\varepsilon \exp(q)$ and $\Gamma = -q$.

\section{Fractional Attractors}
\label{R}

\subsection{Standard Map: Fixed and Periodic Points }
\label{FSMs}

It appears that the dependence 
of the SM's fixed point (0,0) stability properties on the map parameter $K$ 
plays an important role when transition to the FSMs is considered.
(0,0) SM fixed point is stable for $0<K<4$. At $K=4$ it becomes unstable
but period two ($T=2$) antisymmetric trajectory 
\begin{equation} \label{T2point} 
p_{n+1} = -p_n, \    \  x_{n+1} = -x_n
\end{equation}
appears which is stable for $4<K<2 \pi$. At $K=2 \pi$ this T=2 trajectory
becomes unstable but it gives birth to two $T=2$ trajectories 
\cite{Chirikov,LichLib} with 
\begin{equation} \label{T2pointN} 
p_{n+1} = -p_{n}, \    \  x_{n+1} = x_n-\pi.
\end{equation}
These $T=2$ trajectories become unstable when $K \approx 6.59$ 
(see Figure~\ref{figKc}a), 
at the point where $T=4$ stable trajectories are born. 
This period doubling cascade of
bifurcations sequence continues with $T=8$ and $T=16$ stable trajectories 
appearing at $K \approx 6.63$ and $K \approx 6.6344$ correspondingly.  
It continues until at
$K \approx 6.6345$
all periodic points become unstable and corresponding islands disappear.
This scenario of the elliptic-hyperbolic point transitions with the births
of the double periodicity islands inside the original island
has been investigated in \cite{Schmidt} and applied to investigate the SM
stochasticity at low values of the parameter $K<4$. 

\subsection{Phase Space at Low $K$ (Stable (0,0) Fixed Point) }

\begin{figure}
\centering
\rotatebox{0}{\includegraphics[width=12. cm]{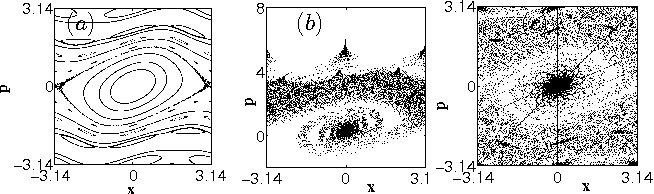}}
\caption{\label{figLowK} (a). 1000 iterations on each of 25 trajectories 
for the SM with $K=0.6$.  The main features are (0,0) fixed point and
T=2 and T=3 trajectories;
(b). 400 iterations on each trajectory with
$p_0=2+0.04i$, $0 \le i < 50$ 
for the FSMRL case $K=0.6$, $\alpha=1.9$. 
Trajectories converging to the fixed point and ASDTs of period 
2 and 3 are present; 
(c).  100 iterations on each FSMC trajectory with $p_0=-3.14+0.0314i$, 
$0 \le i < 200$ for the same case as in Fig.~\ref{figLowK}b 
($K=0.6$, $\alpha=1.9$)
considered on a torus. 
In this case all trajectories converge to the fixed point, period two and 
period three stable attracting points. 
}
\end{figure}

\begin{figure}
\centering
\rotatebox{0}{\includegraphics[width=12. cm]{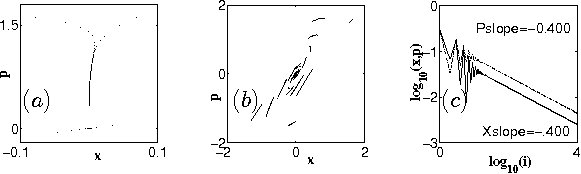}}
\caption{\label{figLowKStab} 
Convergence of trajectories to the fixed point $(0,0)$ at $K<K_{c1}$: 
(a). Two trajectories for the FSMRL with $K=2$,  $\alpha =1.4$, and 
$10^5$ iterations on each trajectory. The bottom one with $p_0=0.3$ is a fast
converging trajectory. The upper trajectory with $p_0=5.3$ is an example
of the FSMRL's ASCT;  
b). Evolution of the FSMC trajectories with 
$p_0=1.6+0.002i$, $0 \le i < 50$ for 
the case $K=3$, $\alpha=1.9$. 
The line segments correspond to the $n$th iteration on the
set of trajectories with close initial conditions; \ \
c). $x$ and $p$ time dependence for the FSMC with $K=2$, $\alpha =1.4$,
$x_0=0$, and $p_0=0.3$.
}
\end{figure}

\begin{figure}
\centering
\rotatebox{0}{\includegraphics[width=12. cm]{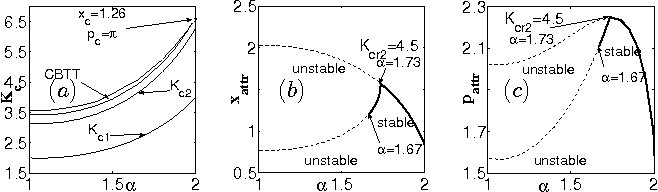}}
\caption{\label{figKc} Critical values in the FSMs' ($\alpha,K$)-space.
The results obtained by the numerical simulations of equations (\ref{Kcr1}), 
(\ref{T2nonASKc}), (\ref{xl1}), (\ref{T2nonASxsol}), (\ref{RLplim}) 
and confirmed by the direct simulations of the FSMs:
(a). The $(0,0)$ fixed point is stable below $K=K_{c1}$ curve. It becomes 
unstable at  $K=K_{c1}$ and gives birth to the antisymmetric $T=2$
sink which is stable at $K_{c1}<K<K_{c2}$. A pair of $T=2$ sinks with
$x_{n+1}=x_n-\pi$,  $p_{n+1}=-p_n$ is stable in the band above $K=K_{c2}$
curve. Cascade of bifurcations type trajectories (CBTTs) are the only
phase space features that appear and exist in the narrow band which ends 
at the cusp on the figure. $(x_c,p_c)$ is the point at which the SM's $T=2$    
elliptic points with $x_{n+1}=x_n-\pi$,  $p_{n+1}=-p_n$ become unstable
and bifurcate (see Sec.~\ref{FSMs});
(b). $\alpha$-dependence of the $T=2$ sinks' x-coordinate for $K=4.5$.
The transition from the antisymmetric sink (upper curve) to $x_{n+1}=x_n-\pi$,  
$p_{n+1}=-p_n$ point takes place at $\alpha=1.73$. Solid lines represent
areas of stability;
(c). $K=4.5$   $\alpha$-dependence of the $T=2$ sinks' p-coordinate.
}
\end{figure}

Stability of the fixed point (0,0) has been considered in
\cite{ETFSM,myFSM}. It has been shown that for fractional values of 
$\alpha$ (we consider only $1<\alpha \le 2$)  this point turns from
elliptic (for  $\alpha = 2$) into a sink which is stable for 
\begin{equation} \label{Kc} 
K<K_{c1}(\alpha)= \frac{2 \Gamma(\alpha)}{V_{\alpha l}},
\end{equation}
where
\begin{equation} \label{Val} 
 V_{\alpha l}  =  \sum_{k=1}^{\infty} (-1)^{k+1} V_{\alpha}^1(k)
\end{equation}
and can be calculated numerically (see Figure~\ref{figKc}a). 
It can be shown that 
$K_{c1}(2) = 4$, which corresponds to the SM case. 
The structure of the FSM phase space
for  $K<K_{c1}$ preserves some features which exist in the $\alpha = 2$ case.  
Namely, stable higher period points, which exist in the SM
case, still exist in the FSM, but they exist in the asymptotic sense and
they transform into sinks or, in the FSMRL case, into attracting slow 
($p_n \sim n^{2-\alpha}$) diverging trajectories (ASDT) 
(see Figure~\ref{figLowK}). 
Pure chaotic trajectories disappear. Each attractor has its own basin of
attraction. The traces of the SM chaotic sea exist in the following sense:
initially close trajectories, which do not start from any basin of 
attraction, may fall into absolutely different attractors. In most 
of the cases attracting points themselves do not belong to 
their own basins of attraction: trajectory which starts from an attracting
point may fall into a  different attractor. The 
FSMRL trajectories which converge to the fixed point follow two different 
routs. Trajectories which start from the basin of attraction converge according
to the power law $x_n \sim n^{-1-\alpha}$  and  $p_n \sim n^{-\alpha}$,
while those starting from the outside of the basin of attraction are
attracting slow converging trajectories (ASCT) 
with $x_n \sim n^{-\alpha}$  and  
$p_n \sim n^{1-\alpha}$. All FSMC trajectories converging to $(0,0)$ 
fixed point follow the slow power law  $x_n \sim n^{1-\alpha}$  and  
$p_n \sim n^{1-\alpha}$ (see Figure~\ref{figLowKStab}). 
This convergence may correspond to the 
recently introduced for the fractional dynamic systems 
notion of the generalized Mittag-Leffler stability 
\cite{MLStab}, for which the power-law stability is a special case.

 \subsection{Phase Space at  $K_{c1}<K<K_{c2}$ (Stable 
$T=2$  Antisymmetric Trajectory) }

\begin{figure}
\centering
\rotatebox{0}{\includegraphics[width=12. cm]{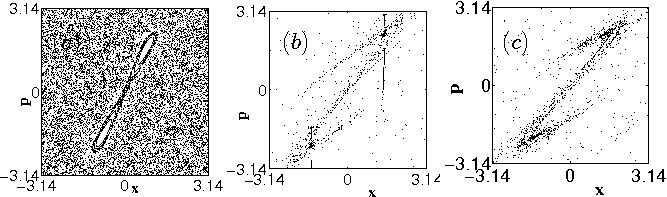}}
\caption{\label{FigT2AS} Stable antisymmetric 
$x_{n+1}=-x_n$,  $p_{n+1}=-p_n$ period   $T=2$ trajectories for
$K=4.5$: (a). 1000 iterations on each of 25 trajectories 
for the SM with $K=4.5$.  The only feature is a system of two
islands associated with the period two elliptic point;
(b). The FSMRL stable $T=2$ antisymmetric sink 
for $\alpha=1.8$.  500 iterations on each of 25 trajectories: 
$p_0=0.0001+0.08i$, $0 \le i <25$. Slow and fast converging trajectories;
(c). The FSMC stable $T=2$ antisymmetric sink 
for $\alpha=1.8$. 1000 iterations on each of 10 trajectories: 
$p_0=-3.1415+0.628i$, $0 \le i <10$. 
}
\end{figure}

Numerical simulations confirm 
that, as in the SM case, in the FSMs  a period 
$T=2$ antisymmetric trajectory (sink) exists (asymptotically) 
for  $K_{1}(\alpha)<K<K_{2}(\alpha)$ 
($4<K<2\pi$ in the SM case) with
\begin{equation} \label{T2AS} 
p_{n+1} = -p_n, \    \  x_{n+1} = -x_n.
\end{equation}
Asymptotic existence and stability
of this sink is a result of the gradual transformation of the SM's
elliptic point with the decrease in the
order of derivative from  $\alpha =2$  
(see Figure~\ref{FigT2AS}). As in the fixed point
case, there are two types of convergence of the trajectories to the
FSMRL $T=2$ sink: fast with 
\begin{equation} \label{FastConv} 
\delta x_n \sim n^{-1-\alpha}, \  \ \delta p_n
\sim n^{-\alpha}
\end{equation}
and slow with  
\begin{equation} \label{SlowConv}
\delta x_n \sim n^{-\alpha}, \  \  \delta p_n;
\sim n^{1-\alpha}
\end{equation} 
and 
\begin{equation} \label{CaputoConv}
\delta x_n \sim n^{1-\alpha}, \  \  \delta p_n
\sim n^{1-\alpha} 
\end{equation}
convergence in the FSMC case \cite{myFSM}.

Let's consider the FSMRL and assume the 
asymptotic existence of the $T=2$ sink 
$(x_l, p_l)$ 
and $(-x_l, -p_l)$.
In this case the limit $n \rightarrow \infty$ in  (\ref{FSMRLp}) and (\ref{FSMRLx}) gives
\begin{equation} \label{RLplim} 
p_l = \frac{K}{2} \sin(x_l),
\end{equation}
\begin{equation} \label{RLxlim} 
x_{l} = \frac{K}{2 \Gamma(\alpha)} \sin(x_l) \sum_{k=1}^{\infty} (-1)^{k+1}
V_{\alpha}^1(k).
\end{equation}
The equation for $x_l$ takes the form
\begin{equation}\label{xl1} 
x_l = \frac{K}{2 \Gamma(\alpha)} V_{\alpha l} \sin(x_l),
\end{equation}
from which the condition of the existence of the $T=2$ sink is   
\begin{equation} \label{Kcr1} 
K > K_{c1}(\alpha) = \frac{2 \Gamma(\alpha)}{V_{\alpha l}}.
\end{equation}
From (\ref{Kcr1}) it follows that  $K_1(\alpha) = K_{c1}(\alpha)$.

\begin{itemize}
\item {\bf Remark 1:} The $T=2$ sink exists only in the
  asymptotic sense. On a trajectory which starts from point 
$(x_0,p_0)$=$(x_l,p_l)$  $x_2 \ne x_l$ and $p_2 \ne p_l$.   
\item {\bf Remark 2:} In spite of the fact that in the derivation
of (\ref{RLplim}) and (\ref{RLxlim}) we used  the limit $n \rightarrow
\infty$, computer simulations demonstrate convergence of all trajectories
to the limiting values
in the very good agreement with (\ref{FastConv}), (\ref{SlowConv}), and
(\ref{CaputoConv}).  
\item {\bf Remark 3:} Direct computations show that  
$K_1(\alpha) = K_{c1}(\alpha)$ is valid for the FSMC too. 
 \end{itemize}

\subsection{Phase Space at  $K>K_{c2}$}

\begin{figure}
\centering
\rotatebox{0}{\includegraphics[width=12. cm]{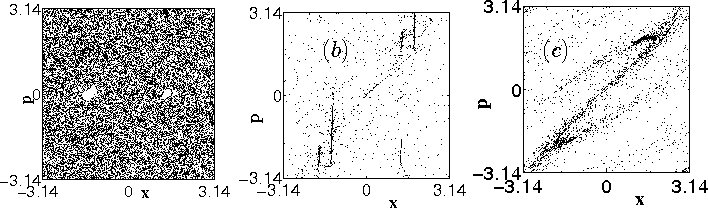}}
\caption{\label{FigT2nonAS} Stable  
$x_{n+1}=x_n-\pi$,  $p_{n+1}=-p_n$ period   $T=2$ trajectories for
$K>K_{c2}$: (a). 500 iterations on each of 50 trajectories 
for the SM with $K=6.4$.  The main features are two accelerator mode 
sticky islands around points $(-1.379,0)$ and $(1.379,0)$ which define 
the dynamics. Additional features - dark spots at the top and the bottom
of the figure (which are clear on a zoom) - 
two systems of $T=2$ tiny islands 
associated with two  $T=2$ elliptic points:  $(1.379,\pi)$,
$(1.379-\pi,-\pi)$ and $(\pi-1.379,\pi)$, $(-1.379,-\pi)$;
(b). Two FSMRL's stable $T=2$ sinks for $K=4.5$, $\alpha=1.71$. 
500 iterations on each of 25 trajectories: 
$p_0=0.0001+0.08i$, $0 \le i <25$;
(c). Two FSMC's stable $T=2$ sinks 
for  $K=4.5$, $\alpha=1.71$. 1000 iterations on each of 10 trajectories: 
$p_0=-3.1415+0.628i$, $0 \le i <10$. 
}
\end{figure}

The SM's $T=2$ antisymmetric trajectory 
becomes unstable when $K=2\pi$, at the point
in phase space where a pair of $T=2$ trajectories with 
$x_{n+1}=x_n-\pi$,  $p_{n+1}=-p_n$ appears. Numerical simulations show 
(see Figure~\ref{FigT2nonAS}) that the FSMs demonstrate similar behavior.  
Let's assume that the FSMRL equations (\ref{FSMRLp}) and (\ref{FSMRLx}) 
have an asymptotic solution
\begin{equation} \label{T2nonAS} 
p_{n} = (-1)^np_l, \    \  x_{n} = x_l-\frac{\pi}{2}[1-(-1)^n].
\end{equation}
Then it follows  from (\ref{FSMRLp}) that relationship (\ref{RLplim}) 
$p_l = K/2 \sin(x_l)$
holds in this case too.
Simulations similar to those presented in Figure~\ref{figLowKStab}b 
\cite{ChaosCBTT} show that for $K>K_{c2}$ (see Figure~\ref{figKc}a) the FSMRL 
has an asymptotic solution
\begin{equation} \label{T2nonLimAS} 
p_{n} = (-1)^np_l+An^{1-\alpha}
\end{equation}
with the same $A$ for both even and odd values of $n$.
Substituting (\ref{T2nonLimAS}) in (\ref{FSMRLx}) and considering  
limit $n \rightarrow \infty$ one can derive (see \cite{ChaosCBTT})
\begin{equation} \label{T2nonASxsol} 
\sin(x_l)= \frac{\pi \Gamma(\alpha)}{K V_{\alpha l}}, 
\end{equation}
which has solutions for 
\begin{equation} \label{T2nonASKc} 
K>K_{c2}= \frac{\pi \Gamma(\alpha)}{V_{\alpha l}} 
\end{equation}
(see Figure~\ref{figKc}) and 
the value of $A$ can also be calculated:
\begin{equation} \label{T2nonASA} 
A= \frac{2 x_l-\pi}{2 \Gamma(2-\alpha)}. 
\end{equation}
These results are in a good agreement with the direct FSMRL numerical
simulations and hold for the FSMC.

\subsection{Cascade of Bifurcations Type Trajectories}

\begin{figure}
\centering
\rotatebox{0}{\includegraphics[width=12. cm]{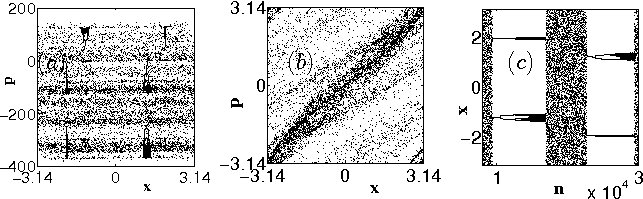}}
\caption{\label{FigCBTT} Cascade of bifurcations type trajectories (CBTT)
from iterations on a single trajectory with $K=4.5$, $\alpha=1.65$,
$x_0=0$, and $p_0=0.3$: 
(a). 25000 iterations for the FSMRL case. 
The trajectory occasionally sticks to one of the cascade of bifurcations 
type trajectories but later always returns to the chaotic sea;
(b). 30000 iterations for the FSMC case.
The CBTTs can hardly be recognized on the full phase portrait but can be seen
on the $x$ of $n$ dependence in Figure~\ref{FigCBTT}c;  
(c). Time dependence of the coordinate $x$ in Figure~\ref{FigCBTT}b.
}
\end{figure}

In the SM further increase in $K$  at $K  \approx 6.59$ causes  
an elliptic-hyperbolic point transition when   
$T=2$ points become unstable and stable $T=4$ elliptic points appear. 
A period doubling cascade of bifurcations leads to the disappearance 
of the corresponding islands of stability in the chaotic sea at 
$K \approx 6.6344$. The cusp in Figure~\ref{figKc}a points to approximately
this spot ($\alpha=2$, $K \approx  6.63$). Inside of the band leading to 
the cusp a new type of the FSM's attractors appears: 
cascade of bifurcations type trajectories (CBTT) (see Figure~\ref{FigCBTT}).
In CBTTs period doubling cascade of bifurcations occurs on a single
trajectory without any change in the map parameter.
In some cases  CBTTs behave like sticky islands of the Hamiltonian
dynamics: occasionally a trajectory enters a CBTT and then leaves it
entering the chaotic sea.
Near the cusp CBTTs are barely distinguishable - the relative time
trajectories spend in  CBTTs is small. With decrease in $\alpha$
the relative time trajectories spend in CBTTs increases and 
at $\alpha$ close to one a trajectory enters a CBTT after a few iterations
and stays there over the longest computational time we were able to
run our codes - 500000 iterations. A typical FSMC's CBTT of this kind is
presented in the Figure~\ref{FigMix}c.   

CBTTs, easily recognizable in the phase space of the
FSMRL (Figure~\ref{FigCBTT}a), often almost
impossible to distinguish in the phase space of the FSMC
(Figure~\ref{FigCBTT}b). But they 
clearly reveal themselves on a coordinate versus time (step of iteration
$n$) plot (Figure~\ref{FigCBTT}c).

\subsection{More FSM's attractors}

\begin{figure}
\centering
\rotatebox{0}{\includegraphics[width=12. cm]{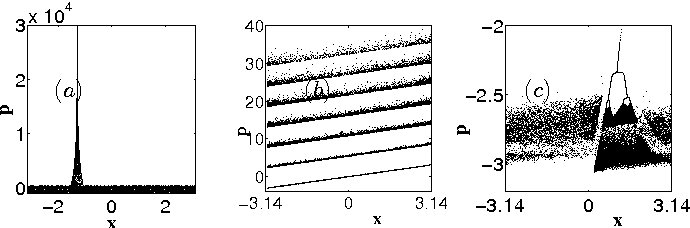}}
\caption{\label{FigMix} 
(a). A single FSMRL ballistic trajectory for $K=1.999$, $\alpha=6.59$;
(b). Seven disjoint chaotic attractors for the case FSMRL with 
$K=4.5$ and  $\alpha=1.02$. 1000 iterations 
on each of 20 trajectories: $x_0=0$ and  $p_0=0.0001+1.65i$, $0 \le i <20$;
(c). 20000 iterations on each of two overlapping independent attractors 
for the FSMC case with $K=4.5$ and $\alpha=1.02$. The CBTT has $p_0=-1.8855$ 
and the chaotic attractor $p_0=-2.5135$ ($x_0=0$).
}
\end{figure}

The main feature of the SM's phase space at $K>2\pi$, which defines dynamics 
and transport and is well investigated (see for example \cite{ZEN}), 
is the presence of the sticky accelerator mode islands. Slight decrease 
in the order of derivative turns  accelerator mode islands into attracting
accelerator mode trajectories (AMT) 
(Figure~\ref{FigMix}a). Depending on the values of $K$ and $\alpha$
trajectories either permanantly sink into this attractor,
or intermittently stick to it and return to the chaotic sea.
AMTs do not exist (we were unable to find them) for $\alpha<1.995$. 
In the ($\alpha,K$)-plot (Figure~\ref{figKc}a) the area of the AMT
existence is a part ($K>2\pi$) of a very narrow strip near $\alpha=2$ line.
In the area of the  $\alpha,K$-plot  above the CBTT strip (we 
considered only $K<7$) different kinds of chaotic attractors (or possibly pure
chaotic behavior) can be found. Two examples are presented in 
Figure~\ref{FigMix}. Figure~\ref{FigMix}c contains two overlapping
attractors one of which is a CBTT. Existence of the overlapping attractors
and intersecting trajectories is a feature of maps with memory which 
is impossible in a regular dynamical systems. The simplest example of 
intersecting trajectories can be constructed by starting one trajectory 
at an arbitrary point $(x_0,p_0)$ and then the second at 
$(x'_0,p'_0)=(x_1,p_1)$.

\subsection{The Fractional Dissipative Standard Map (FDSM)}

Consideration of the FDSM in \cite{TEdisFM} demonstrated that already
a small deviation of $\alpha$ from $2$ leads to significant changes in 
properties of the map. For example, a window of ballistic motion which
exists in Zaslavsky Map near $K=4\pi$ is closing when  $\alpha<1.996$. 
Further decrease in  $\alpha$ at  $K \approx 4\pi$ produces different
kinds of chaotic attractors and sinks \cite{TEdisFM}.

\begin{figure}
\centering
\rotatebox{0}{\includegraphics[width=12. cm]{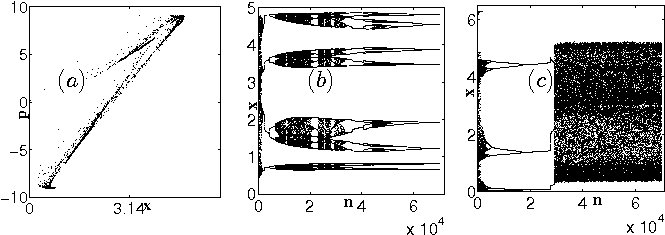}}
\caption{\label{FigFDSM} The FDSM (in all cases $\Gamma=5$ and $\Omega=0$):
(a). Phase space for the case $K=9.1$, $\alpha=1.2$;
(b). 70000 iterations on a single trajectory for the same case as in 
Figure~\ref{FigFDSM}a ($K=9.1$, $\alpha=1.2$). 
A cascade of bifurcations appears  from chaos and
then converges into a periodic trajectory; 
(c). 70000 iterations on a single trajectory for the  case $K=10$,
$\alpha=1.2$. A chaotic attractor 
appears in the FDSM's phase space after a chaotic trajectory 
through an inverse cascade of
bifurcations forms a period $T=3$ trajectory.
}
\end{figure}

Some of the new features we observed, which are presented in 
Figure~\ref{FigFDSM}, include inverse cascade of bifurcations
and trajectories which start as a CBTT and then converge
as an inverse cascade of bifurcations. 
In phase space those trajectories may look like chaotic attractors.


\section{Conclusion}
\label{C}

We studied three different fractional maps  which describe
nonlinear systems with memory under periodic perturbations (kicks in our
case). The types of solutions which we found in all maps include
periodic sinks, attracting slow diverging trajectories,
attracting accelerator mode trajectories, 
chaotic attractors, and cascade of
bifurcations type trajectories. New features discovered include
attractors which overlap, trajectories which intersect, and CBTTs.

The models with a similar behavior may include periodically kicked
media which can be described by FDEs: viscoelastic materials
\cite{MainardiBook2010} or dielectrics \cite{TarDiel}. 
Experiments can be proposed to observe different kinds of solution, 
for example CBTTs, in those media.

Nonlinear models with memory which have chaotic and periodic solutions, 
where cascades of bifurcations considered as a result of a change 
in a system parameter, are used in population biology and 
epidemiology \cite{Hopp1976}. Our calculations show 
that such behavior can be a consequence of the essential internal
properties of the systems with memory and suggest construction of new
fractional models of biological systems. 

Our computational results are based on the runs over more that
$10^5$ periods of perturbations. Due to the integro-differential   
nature of the fractional derivatives, at present time  it is impossible 
to solve FDEs over more than  $10^3$ periods of oscillations. 
Our results suggest a new way of looking for a particular solutions
of the FDEs. For example, we may suggest that CBTTs can be found 
at the values of  system parameters, where in corresponding integer system
a period doubling cascade of bifurcations leads to the disappearance of 
a system of islands.

\subsection*{Acknowledgment}

The authors express their gratitude to the administration of the  
Stern College,  Courant Institute, and DOE Grant 
DE-FG0286ER53223 for the financial support.

\end{document}